\documentclass[prl,twocolumn,superscriptaddress,aps,amsmath,amssymb,nofootinbib]{revtex4}
\usepackage{graphicx}
\usepackage{dcolumn}
\usepackage{bm}
\usepackage{color}

\usepackage[section]{placeins}
\usepackage{graphicx,epsfig}
\usepackage{amsmath}
\usepackage{amsfonts}
\usepackage{braket}
\usepackage{fancyhdr}
\usepackage{hyperref}
\usepackage[utf8]{inputenc}
\usepackage[T1]{fontenc}
\usepackage{amsthm}
\usepackage{hhline}
\usepackage{multirow}
\usepackage[figuresleft]{rotating}
\def\beq{\begin{equation}}
\def\eeq{\end{equation}}
\def\beqa{\begin{eqnarray}}
\def\eeqa{\end{eqnarray}}

\begin{document}



\title{Fermi polaron in low-density spin-polarized neutron matter}

\author{Isaac Vida\~na}
\affiliation{Istituto Nazionale di Fisica Nucleare, Sezione di Catania, Dipartimento di Fisica ``Ettore Majorana'', Universit\`a di Catania, Via Santa Sofia 64, I-95123 Catania, Italy}


\begin{abstract}
We study the properties of a spin-down neutron impurity immersed in a low-density free Fermi gas of spin-up neutrons. In particular, we analyze its energy ($E_\downarrow$), effective mass ($m^*_\downarrow$) and quasiparticle residue ($Z_\downarrow$). Results are compared with those of state-of-the-art quantum Monte Carlo calculations of the attractive Fermi polaron realized in ultracold atomic gases experiments, and with those of previous studies of the neutron polaron. Calculations are performed within the Brueckner--Hartree--Fock approach using the chiral two-body nucleon-nucleon interaction of Entem and Machleidt at N$^3$LO with a 500 MeV cut-off and the Argonne V18 phenomenological potential. Only contributions from the  $^1S_0$ partial wave, which is the dominant one in the low-density region considered, are included. Contributions from three-nucleon forces are expected to be irrelevant at these densities and, therefore, are neglected in the calculation. Our results show that for Fermi momenta between $\sim 0.25$ and $\sim 0.45$ fm$^{-1}$  the energy, effective mass and quasiparticle residue of the impurity vary only slightly, respectively, in the ranges $-0.604\,E_F < E_\downarrow < -0.635\,E_F $, $1.300\,m < m^*_\downarrow < 1.085\, m$ and $0.741 <Z_\downarrow< 0.836$ in the case of the chiral interaction, and $-0.621\,E_F < E_\downarrow < -0.643\,E_F $, $1.310\,m < m^*_\downarrow < 1.089\, m$ and $0.739 <Z_\downarrow< 0.832$ when using the Argonne V18 potential.
These results  are compatible with those derived from ultracold atoms and show that a spin-down neutron impurity in a  free Fermi gas of spin-up neutrons with a Fermi momentum in the range $0.25\lesssim k_F \lesssim 0.45$ fm$^{-1}$  exhibits properties very similar to those of an attractive Fermi polaron in the unitary limit.
\end{abstract}


\maketitle


The idea of the polaron, a quasiparticle arising from the dressing of an impurity strongly coupled to an environment or a bath, was introduced in the pioneering works of Landau \cite{Landau33} and Pekar \cite{Pekar46} to describe the properties of conduction electrons in a dielectric crystal, and it was further elaborated by Fr\"{o}hlich \cite{Frohlich54} and Feynman \cite{Feynman55} who considered the ionic crystal or polar semiconductor as a phonon bath. Since this initial introduction of  the electron-phonon polaron concept in solid-state physics, the polaron idea has been extended and generalized to other areas of physics with applications in field-effect transistors \cite{Hulea06}, high-temperature superconductors \cite{Mott95,Muller17},  ultracold atomic gases \cite{Chevy10, Massignan14}, nuclear physics \cite{Kutschera92,Forbes14,Roggero15,Nakano20}, or even general relativity \cite{Bahrami14} and quantum-state reduction \cite{Penrose14}. In particular, experiments with population-imbalanced ultracold atomic gases, where the minority atomic species plays the role of the impurity and the majority one that of the environment, have allowed the experimental realization of polarons providing an exceptional framework for studying the properties of quantum impurities. The possibility of varying the impurity-medium interaction from a weak to a strong coupling regime by means of Feshbach resonances \cite{Chin10} has permitted to investigate, in a controlled way, how the impurity becomes ``dressed'' by the excitations of the medium. The properties of the polaron depend strongly on the quantum statistics of the majority atomic species. Experiments in which the minority atomic species is immersed  in a bath of atoms of fermionic or bosonic nature have been carried out revealing the existence of attractive and repulsive Fermi \cite{Shin08,Schirotzek09,Nascimbene,Kohstall12,Koschorreck12,Cetina15,Ong15,Cetina16,Scazza17,Yan19,Darkwah19,Sous20} and Bose \cite{Chikkatur00,Tempere09,Catani12,Spethmann12,Scelle13,Pena15,Pena16,Rentrop16,Jorgensen16,Hu16,Levinsen17,Pastukhov18,Guenther18,Pena18,Pena19a,Pena19b,Pena19c,Yan00,Pena00a,Pena00b,Pena00c} polarons. Whereas Fermi polarons are the paradigmatic realization of Landau's fundamental idea of quasiparticle, the description of an impurity atom in a Bose-Einstein condensate can be cast in the form of the Fr\"{o}hlich’s polaron Hamiltonian \cite{Frohlich54}, where the role of the phonons is played by Bogoliubov excitations. The theoretical description of Fermi and Bose polarons has achieved a significant progress thanks to the development and application of several numerical many-body techniques that include among others, the renormalization group theory \cite{Jeckelmann96, Grusdt16,Grusdt17}, exact diagonalization \cite{Fehske07}, field-theoretical diagrammatic approaches \cite{Rath13}, quantum Monte Carlo \cite{Lobo06, Pilati08,Kornilovitch07}, diagrammatic Monte Carlo methods \cite{Mishchenko00,Prokofev08,Hahn18,vanHoucke20} and variational approaches \cite{Li14,Levinsen15,Shchalidova16,Yoshida18,Loon18,Drescher19}. 

Nowadays most of the interest is focused on the Bose polaron, although the Fermi polaron is still an exciting area of research. In this work, we study the properties of a spin-down ($\downarrow$) neutron impurity immersed in a low-density free Fermi gas of spin-up ($\uparrow$) neutrons and show that it behaves as an attractive Fermi polaron in the unitary limit, {\it i.e.,} the limit of infinite (negative) $S$-wave scattering length of a gas of fermions at vanishing small density. Despite the fact that the neutron-neutron scattering length of the $^1S_0$ partial wave is extremely large ($a_S=-18.5$ fm), low-density neutron matter actually never reaches the unitary limit, although it shows properties ``close''  to it in the range $r_e<n^{-1/3}<|a_S|$, where $r_e=2.75$ fm is the effective range of the $^1S_0$ neutron-neutron interaction and $n^{-1/3}$ is the average interparticle spacing, with $n$ being the density of neutron matter \cite{Baldo08,Gezerlis10,Gezerlis11,Gezerlis12,Kruger15}. A few years ago, Forbes {\it et al.} \cite{Forbes14} extended the idea of the polaron to a system of strongly interacting neutrons. These authors studied the energy of the neutron polaron performing quantum Monte Carlo calculations using two nucleon-nucleon (NN) interactions, the Argonne V18 potential \cite{Wiringa95} and a modified P\"{o}schl--Teller potential \cite{Poschl33}, and compared their results with those obtained from effective field theory calculations that included contributions beyond the effective range of the impurity-fermion interaction. The effective mass of the impurity was also determined by these authors, finding values compatible with those of the Fermi polaron in the unitary limit. Furthermore, these authors used the neutron polaron energy to constrain the time-odd part of nuclear energy density functionals (EDFs) in order to obtain better descriptions of polarized nuclear systems. Similarly, Roggero {\it et al.} \cite{Roggero15} used the  quantum Monte Carlo method with chiral NN interactions to analyze the energy of a proton impurity in low-density neutron matter finding that, for a wide range of densities, the behavior of the proton impurity is similar to that of a polaron in a fully polarized unitary Fermi gas. The authors of this work, as those of Ref.\ \cite{Forbes14}, employed also their results to impose tight constraints on the time-odd components of Skyrme-type  nuclear EDFs. In the present work, in addition to the energy, we also analyze the effective mass and the quasiparticle residue (or $Z$ factor) of the $\downarrow$ neutron impurity, and we compare the results obtained with state-of-the-art quantum Monte Carlo calculations of the attractive Fermi polaron realized in ultracold atomic gases experiments. 

Our calculation starts with the construction of the  $\langle \vec k_\downarrow \vec k_\uparrow |G(\omega )|\vec k_\downarrow \vec k_\uparrow \rangle$ $G$-matrix elements describing the in-medium interaction of the $\downarrow$ neutron impurity with one of the $\uparrow$ neutrons of the free Fermi gas. To such end, using the chiral NN interaction of Entem and Machleidt at N$^3$LO with a 500 MeV cut-off  (hereafter referred to simply as EM500) \cite{Entem03} and the Argonne V18 phenomenological potential \cite{Wiringa95} as bare interactions between the $\downarrow$ and $\uparrow$ neutrons, we solve the coupled-channel Bethe--Goldstone integral equation, 
\begin{widetext}
\begin{equation}
\langle \vec k_{\sigma_1} \vec k_{\sigma_2} |G(\omega)|\vec k_{\sigma_3} \vec k_{\sigma_4} \rangle =
\langle \vec k_{\sigma_1} \vec k_{\sigma_2} |V|\vec k_{\sigma_3} \vec k_{\sigma_4} \rangle
+\sum_{\sigma_i\sigma_j}
\frac{\langle \vec k_{\sigma_1} \vec k_{\sigma_2} |V|\vec k_{\sigma_i} \vec k_{\sigma_j} \rangle
 \langle \vec k_{\sigma_i}\vec k_{\sigma_j}|Q|\vec k_{\sigma_i}\vec k_{\sigma_j}\rangle
 \langle \vec k_{\sigma_i} \vec k_{\sigma_j} |G(\omega)|\vec k_{\sigma_3} \vec k_{\sigma_4} \rangle }
 {\omega-E_{\sigma_i}(\vec k_{\sigma_i})-E_{\sigma_j}(\vec k_{\sigma_j})+i\eta}
 \ ,
\label{eq:bbg}
\end{equation}
\end{widetext}
including only contributions from the  $^1S_0$ partial wave which is the dominating one in the low-density region \cite{Baldo08}. Contributions from three-nucleon forces are expected to be irrelevant at these densities and, therefore, are neglected in the calculation \cite{Hebeler10, Tews13,Kruger13}. In Eq.\ (\ref{eq:bbg}), $V$ is the bare NN interaction ({\it i.e.,} the EM500 or the Argonne V18),  $Q$ is the Pauli operator allowing only intermediate states compatible with the Pauli principle, and $\omega$ is the starting energy, defined as the sum of the initial non-relativistic energies of the interacting  $\downarrow$ and $\uparrow$ neutrons.  Whereas the energy of a $\uparrow$ neutron of the free Fermi gas is just kinetic that of the $\downarrow$ neutron impurity is $E_\downarrow(\vec k_\downarrow)=\hbar^2k^2_\downarrow/2m+\mbox{Re}[U_\downarrow(\vec k_\downarrow)]$ where
\begin{equation}
U_{\downarrow}(\vec k_\downarrow)=\sum_{|\vec k_\uparrow|\leq k_{F_\uparrow}}
\langle \vec k_\downarrow \vec k_\uparrow |G(\omega=E_\downarrow(\vec k_\downarrow)+\frac{\hbar^2 k^2_\uparrow}{2m})|\vec k_\downarrow \vec k_\uparrow \rangle 
\label{eq:spp}
\end{equation}
is the on-shell Brueckner--Hartree--Fock (BHF) $\downarrow$ neutron potential denoting the mean field ``felt'' by the  impurity due to its interaction with the free Fermi gas. We note that the continuous prescription is adopted when solving the Bethe--Goldstone equation. We note also that Eqs.\ (\ref{eq:bbg}) and (\ref{eq:spp}) are self-consistently solved.

\begin{figure}[h!]
\centering
\includegraphics[width=0.9 \columnwidth]{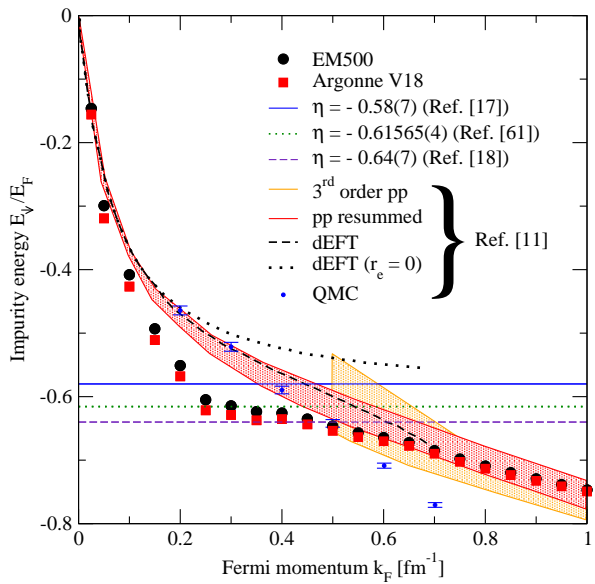}
        \caption{Energy of a $\downarrow$ neutron impurity with $\vec k_\downarrow=\vec 0$ in units of the Fermi energy as a function of the Fermi momentum of the free gas of $\uparrow$ neutrons. Full circles and squares show, respectively,
         the results of our BHF calculation obtained with the EM500 interaction and the Argonne V18 potential. The horizontal lines show the result for the proportionality constant $\eta$ derived experimentally in Refs.\ \cite{Shin08,Schirotzek09} (solid and dashed lines) 
        and in the quantum Monte Carlo calculation of Ref.\ \cite{vanHoucke20} (dotted line). The results obtained by Forbes {\it et al.} in Ref.\ \cite{Forbes14} are also reported for comparison.}
\label{fig:fig1}
\end{figure}

\begin{figure}[h!]
\centering
\includegraphics[width=0.9 \columnwidth]{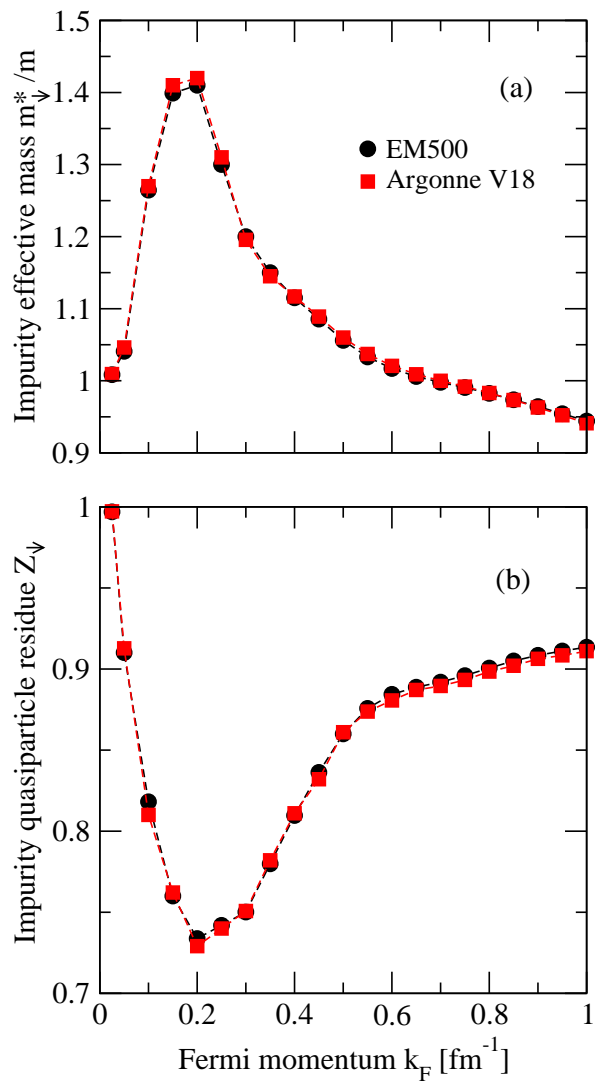}
        \caption{Effective mass (panel a) and quasiparticle residue (panel b) of a $\downarrow$ neutron impurity with $\vec k_{\downarrow}=\vec 0$ as a function of the Fermi momentum of the free gas of $\uparrow$ neutrons. Results are shown for the two NN interaction considered.}
\label{fig:fig2}
\end{figure}

The energy of a Fermi polaron, defined as the change in the energy of the non-interacting Fermi gas when an impurity of zero momentum is added to it, shows in the unitary limit a universal behavior regardless the nature of the impurity, the bath and the details of their mutual interactions, being simply proportional to the Fermi energy, $E_{pol}=\eta E_F$ \cite{Chevy06}. State-of-the-art quantum Monte Carlo calculations of Van Houcke {\it et al.}\ \cite{vanHoucke20} predict a value of the proportionality constant $\eta=-0.61565(4)$, in very good agreement with the values $\eta=-0.58(5)$ \cite{Shin08} and $\eta=-0.64(7)$ \cite{Schirotzek09} extracted from experiments with spin-polarized  $^6$Li atoms with resonant interactions. In  Fig.\ \ref{fig:fig1} we present the energy of a $\downarrow$ neutron impurity with $\vec k_\downarrow=\vec 0$ in units of the Fermi energy as a function of the Fermi momentum of the free gas of $\uparrow$ neutrons. Full circles and squares display, respectively, the results obtained with the EM500 interaction and the Argonne V18 potential. The horizontal lines show the result for the proportionality constant $\eta$ derived experimentally in Refs.\ \cite{Shin08,Schirotzek09} (solid and dashed lines) and in the calculation of  Van Houcke {\it et al.}\ \cite{vanHoucke20} (dotted line).  The results obtained by Forbes {\it et al.} in Ref.\ \cite{Forbes14} are also shown for comparison.  These results were obtained from chiral NN interactions at N$^3$LO at the level of third-order particle-particle (pp) ladder contributions (orange band) and resumming pp ladders (red band); from a difermion effective field theory (dEFT) with (dashed line) and without (dotted line) an effective range of the interaction; and with a quantum Monte Carlo calculation with 33+1 particles and $S$-wave interactions (points with error bars). The reader is referred to Ref.\ \cite{Forbes14} for details of these calculations. We note first that our results are very similar for the two NN interactions employed. This indicates, as expected, that  the details of the NN interaction are irrelevant for neutron matter at very low densities. We also notice that for Fermi momenta between $\sim 0.25$ and $\sim 0.45$ fm$^{-1}$ the ratio $E_\downarrow/E_F$ exhibits a kind of plateau where it varies only slightly between $-0.604$ and $-0.635$ in the case of the EM500 interaction, and between $-0.621$ and $-0.643$ when we use the Argonne V18 potential. These values are compatible with those of the proportionality constant $\eta$ derived in Refs.\ \cite{Shin08,Schirotzek09} and Ref.\ \cite{vanHoucke20}. Furthermore, in this region of Fermi momenta the condition $r_e<n^{-1/3}<|a_S|$ is fulfilled and, hence, low-density neutron matter shows properties ``close'' to those of a unitary Fermi 
gas \cite{Baldo08,Gezerlis10,Gezerlis11,Gezerlis12,Kruger15}. In particular, since for $0.25\lesssim k_F \lesssim 0.45$ fm$^{-1}$ the ratio $E_\downarrow/E_F$ is almost constant, and its value is compatible with those of $\eta$ obtained in Refs.\ \cite{Shin08,Schirotzek09,vanHoucke20}, it is reasonable to conclude that a $\downarrow$ neutron impurity in a free Fermi gas of $\uparrow$ neutrons with a Fermi momentum within this range of values presents a behavior similar to that of an attractive Fermi polaron in the unitary limit. To further confirm this conclusion, in the next, we analyze also the effective mass and the quasiparticle residue of a $\downarrow$ neutron impurity with zero momentum. However, before doing it, we would like to note first that our results for the ratio $E_\downarrow/E_F$ are compatible with those of Forbes {\it et al.} \cite{Forbes14} only at very low densities ($k_F\lesssim 0.1$ fm$^{-1}$) and for values of the Fermi momentum above $\sim 0.5$ fm$^{-1}$, whereas they show discrepancies in the region $0.1\lesssim k_F \lesssim 0.5$ fm$^{-1}$. In particular, we notice that, except for the dEFT calculation with a zero effective range which approaches the value of $\eta$ at the unitary limit with increasing $k_F$, the results of Forbes and collaborators do not show any plateau similar to the one found in our case. A possible explanation of the differences found between our results and those of Ref.\ \cite{Forbes14} could be that in our calculation of the polaron energy we consider, through the $G$-matrix, only the contribution from the rescattering of the $\downarrow$ neutron impurity with the $\uparrow$ neutrons of the free gas and we ignore, for instance, the contribution from the coupling of the impurity with the particle-hole excitations of the gas. In other words, in our calculation only short-range correlations are taken into account, whereas long-range ones are ignored. However, although this is a plausible reason, a deep comparative analysis would be required for a better understanding of the origin of these differences.

Results for the effective mass and the quasiparticle residue of a $\downarrow$ neutron impurity with zero momentum are shown, respectively, in panels a and b of Fig.\ \ref{fig:fig2} as a function of the Fermi momentum of the $\uparrow$ neutron free Fermi gas for the two NN interactions considered.

The effective mass of a $\downarrow$ neutron impurity with zero momentum, $m^*_\downarrow$, can be extracted by assuming that its energy is quadratic for  low values of its momentum $\vec k_{\downarrow}$, and fitting this parabolic energy  to the calculated BHF one, $E_\downarrow(\vec k_{\downarrow})$. The quasiparticle residue is defined as
\begin{equation}
Z_\downarrow=\left(1-\frac{\partial U_\downarrow(\vec k_{\downarrow}=\vec 0,\tilde{E}_{\downarrow})}{\partial \tilde{E}_{\downarrow}}\right)^{-1}_{\tilde{E}_\downarrow=U_\downarrow(\vec k_{\downarrow}=\vec 0)} \ ,
\label{eq:zfact}
\end{equation}
where $U_\downarrow(\vec k_{\downarrow},\tilde{E}_{\downarrow})$ is the off-shell BHF $\downarrow$ neutron impurity potential obtained by integrating off-shell $G$-matrix elements. These are determined by solving the Bethe--Goldstone equation for values of the starting energy  $\omega=\tilde{E}_\downarrow+\hbar^2k^2_\uparrow/2m$ where the single-particle energy $\tilde{E}_\downarrow$ of the $\downarrow$ neutron impurity is not connected with its corresponding momentum $\vec k_\downarrow$ through an energy-momentum dispersion relation. The interested reader can find further details on the general off-shell properties of the nucleon single-particle potential in Ref.\ \cite{Baldo92}. 

The quasiparticle residue $Z_\downarrow$ gives a measurement of the importance of the correlations. The smaller its value is, the more important are the correlations. We find, as it is seen in Fig.\ \ref{fig:fig2}, that $m^*_\downarrow$ ($Z_\downarrow$) increases (decreases) initially, reaches a maximum (minimum) at 
$k_F\sim 0.2$ fm$^{-1}$ and then it decreases (increases) at higher values of $k_F$. To understand better the dependence of $m^*_{\downarrow}$ and $Z_\downarrow$ with $k_F$, we note that, in the Fermi momentum region where $m^*_{\downarrow}$ and $Z_\downarrow$ present their respective maximum and minimum, the average interparticle spacing $n^{-1/3}$ is of the order of the $^1S_0$ neutron-neutron scattering length, {\it i.e.,} $n^{1/3}|a|\sim 1$. Therefore, we can venture to say that this region establishes the border between a less correlated and a more correlated regime of the system. In fact, note that  the values of  $Z_\downarrow$ are in general larger in the $k_F$ region from $0$ to $0.2$ fm$^{-1}$ than for $k_F \gtrsim 0.2$ fm$^{-1}$, indicating that in this region correlations are less important, and that here the $\downarrow$ neutron impurity propagates more freely in the $\uparrow$ neutron gas. Furthermore, in the region of Fermi momenta between $\sim 0.25$ and $\sim 0.45$ fm$^{-1}$ where, as it was said before, low-density neutron matter can be considered  ``close'' to the unitary limit, the effective mass and the quasiparticle residue vary, respectively, in the ranges  $1.300\,m < m^*_\downarrow < 1.085\, m$ and $0.741 <Z_\downarrow< 0.836$ when using the EM500 interaction, and $1.310\,m < m^*_\downarrow < 1.089\, m$ and $0.739 <Z_\downarrow< 0.832$ for the Argonne V18 potential. We note that the range of variation found for both quantities is compatible with the results for the effective mass
of the full-many body analysis of Combescot and Giraud \cite{Combescot08} who found $m^*_\downarrow=1.197\,m$, and those for the quasiparticle residue of the diagrammatic Monte Carlo method employed by Vlietinck {\it et al.} \cite{Vlietinck13} who obtained the value  $Z_\downarrow=0.759$. This confirms once more the Fermi polaron behavior exhibited by the $\downarrow$ neutron impurity in a free Fermi gas of $\downarrow$ neutrons with a Fermi momentum  between $\sim 0.25$ and $\sim 0.45$ fm$^{-1}$.

Summarizing, in this work we have analyzed the properties (energy $E_\downarrow$, effective mass $m^*_\downarrow$ and quasiparticle residue $Z_\downarrow$) of a $\downarrow$ neutron impurity in a low-density free Fermi gas of $\uparrow$ neutrons. Calculations have been done within the BHF approach using the chiral two-body NN force of Entem and Machleidt at N$^{3}$LO with a 500 MeV cut-off and the Argonne V18 phenomenological potential. Only contributions from the $^1S_0$ partial wave have been included. Our results  have shown that  a $\downarrow$ neutron impurity in a free Fermi gas of $\uparrow$ neutrons with Fermi momentum in the range $0.25\lesssim k_F \lesssim 0.45$ fm$^{-1}$ presents properties very similar to those of an attractive Fermi polaron in the unitary limit.

\vspace{0.25cm}
The author is very grateful to Artur Polls, who unfortunately could not see this work finished, for the many interesting and stimulating conversations they had on this topic.
 The author is also very grateful to Marcello Baldo, Silvia Chiacchiera and Albert Feijoo for their comments and careful reading of the manuscript. This work has been supported by the COST Action CA16214,  ``PHAROS: The multi-messenger physics and astrophysics of compact stars''.



\end{document}